\documentclass{emulateapj}

\newcommand{\hi}{H$\;${\small\rm I}\relax}

\newcommand{\cii}{C$\;${\small\rm II}\relax}
\newcommand{\ciii}{C$\;${\small\rm III}\relax}
\newcommand{\silii}{Si$\;${\small\rm II}\relax}
\newcommand{\siliii}{Si$\;${\small\rm III}\relax}
\newcommand{\siliv}{Si$\;${\small\rm IV}\relax}
\newcommand{\nii}{N$\;${\small\rm II}\relax}

\newcommand{\oii}{O$\;${\small\rm II}\relax}
\newcommand{\oiii}{O$\;${\small\rm III}\relax}
\newcommand{\ovi}{O$\;${\small\rm VI}\relax}

\newcommand{\mgii}{Mg$\;${\small\rm II}\relax}

\newcommand{\kms}{km~s$^{-1}$\relax}

\newcommand{\zone}{0.27395}
\newcommand{\ztwo}{0.27416}
\newcommand{\zgone}{0.27406}

\newcommand{\metal}{-0.20 \pm 0.15}

\slugcomment{Accepted for publication in ApJ}
\shortauthors{Ribaudo et al.}
\shorttitle{LLSs and Infall}

\begin{document}

\title{Evidence for Cold Accretion: Primitive Gas Flowing 
onto a Galaxy at $z\sim0.274$\altaffilmark{1}}
\author{ Joseph Ribaudo\altaffilmark{2}\altaffilmark{,}\altaffilmark{3}, 
Nicolas Lehner\altaffilmark{2}, J. Christopher Howk\altaffilmark{2}, 
Jessica K. Werk\altaffilmark{4}, Todd M. Tripp\altaffilmark{5},
J. Xavier Prochaska\altaffilmark{4}, Joseph D. Meiring\altaffilmark{5}, 
\& Jason Tumlinson\altaffilmark{6}}
\altaffiltext{1}{Based on observations with the NASA/ESA Hubble Space Telescope 
obtained at the Space Telescope Science Institute, which is operated by the Association 
of Universities for Research in Astronomy, Incorporated, under NASA contract NAS5-26555.}
\altaffiltext{2}{Department of Physics, University of Notre Dame,  Notre Dame, IN 46556}
\altaffiltext{3}{Current Address: Department of Physics, Utica College, Utica, NY, 13502}
\altaffiltext{4}{UCO/Lick Observatory, University of California, Santa Cruz, Santa Cruz, CA 95064}
\altaffiltext{5}{Department of Astronomy, University of Massachusetts, Amherst, MA 01003}
\altaffiltext{6}{Space Telescope Science Institute, Baltimore, MD 21218}


\begin{abstract}

We present UV and optical observations from the Cosmic Origins
  Spectrograph on the \textit{Hubble Space Telescope} and Keck of a
  $z= \zone$ Lyman limit system (LLS) seen in absorption against the
  QSO PG1630+377. We detect \hi\ absorption with $\log N({\rm
    H\,I})=17.06\pm0.05$ as well as \mgii, \ciii, \siliii, and \ovi\
  in this system. The column densities are readily explained if this
  is a multi-phase system, with the intermediate and low ions arising
  in a very low metallicity ($[\rm Mg/\rm H] =-1.71 \pm 0.06$)
  photoionized gas. We identify via Keck spectroscopy and Large
  Binocular Telescope imaging a 0.3 $L_*$ star-forming galaxy
  projected 37 kpc from the QSO at nearly identical redshift
  ($z=\zgone, \Delta v = -26$ \kms) with near solar metallicity ($[\rm
  O/\rm H]=\metal$).  The presence of very low metallicity gas in the
  proximity of a near-solar metallicity, sub-$L_*$ galaxy strongly
  suggests that the LLS probes gas infalling onto the galaxy.  A
  search of the literature reveals that such low metallicity LLSs are not
  uncommon.  We found that 50\% (4/8) of the well-studied $z\la1$ LLSs
  have metallicities similar to the present system and show sub-$L_*$
  galaxies with $\rho < 100$ kpc in those fields where redshifts
  have been surveyed.  We
  argue that the properties of these primitive LLSs and their host
  galaxies are consistent with those of cold mode accretion streams
  seen in galaxy simulations.

\end{abstract}


\section{Introduction}
\label{sec:intro}

Galaxies are predicted to acquire the majority of their baryons
through cold,\footnote{Cold here implies the gas is not heated to the
  virial temperature; however, the temperature of the gas ($\sim 10^4$
  K) is well above that of the cold material in the disk of the
  galaxy.  } filamentary streams that penetrate deep inside the dark
matter halos without shock-heating to the virial temperature
\citep{bd03,keres05}.  This so-called cold mode accretion (CMA) may be
the primary mechanism by which galaxies acquire material needed for
star formation from the intergalactic medium
(IGM)\citep[][]{bauermeister10}.  Such streams are predicted to have
low temperatures ($T\sim10^4$ K), low metallicities (averaging
$\langle Z \rangle \sim 0.001-0.01 Z_{\odot}$ depending on the
simulation), and to be predominantly ionized
\citep{fumagalli11,fgk11}.  CMA models predict such streams will
remain cold only for galaxies below a threshold mass $M_{\rm
  Halo}\lesssim10^{12} M_\odot$ \citep[or $M_\star\sim5\times10^{10}
M_\odot$;][]{keres05,stewart10}.  Estimates of the covering factor of
cold streams within the virial radius for such galaxies range from
$5\%$ up to $40\%$, depending on redshift and model parameters.
Despite the fundamental role CMA may play in galaxy formation and
evolution, observations have provided little direct evidence for its
existence.

QSO absorption lines can probe the circumgalactic medium of foreground
galaxies, with intermediate \hi\ column density systems ($15.5 \leq
\log N(\mbox{\hi}) \leq 19.0$) being particularly promising tracers of
cold accretion streams \citep[e.g.,][]{fumagalli11,fgk11}.  Among
these, LLSs with $\log N({\rm H\,I}) \geq 16.5$ are both readily
identifiable and often allow for straightforward \hi\ column density
measurements due to the flux discontinuity they cause at the Lyman
limit (912 \AA\ in the rest frame).  The use of LLSs to probe infalling
and outflowing matter near galaxies has the advantage that they are
selected in a metallicity-independent manner.  Thus, unlike searches
for \mgii\ and other metal line-selected absorbers, LLS searches are
not biased in favor of either metal-enriched winds or infalling matter in galaxy
halos.  Furthermore, when the strength of the Lyman break is not too
strong (i.e., $\tau \lesssim 3$), the \hi\ column density of the
system can be measured, from which the metallicity and, ultimately,
the origins of the absorbing gas can be determined.  These features
circumvent some of the problems that currently cause disagreements in
the interpretation of the strong \mgii-selected
absorbers, for which some studies suggest the absorbers trace
outflowing material while others suggest they may represent infalling
material \citep{bouche07,mc09,chen10,bc11,kacprzak11}.


\begin{figure*}[t]
\epsscale{1.0}
\plotone{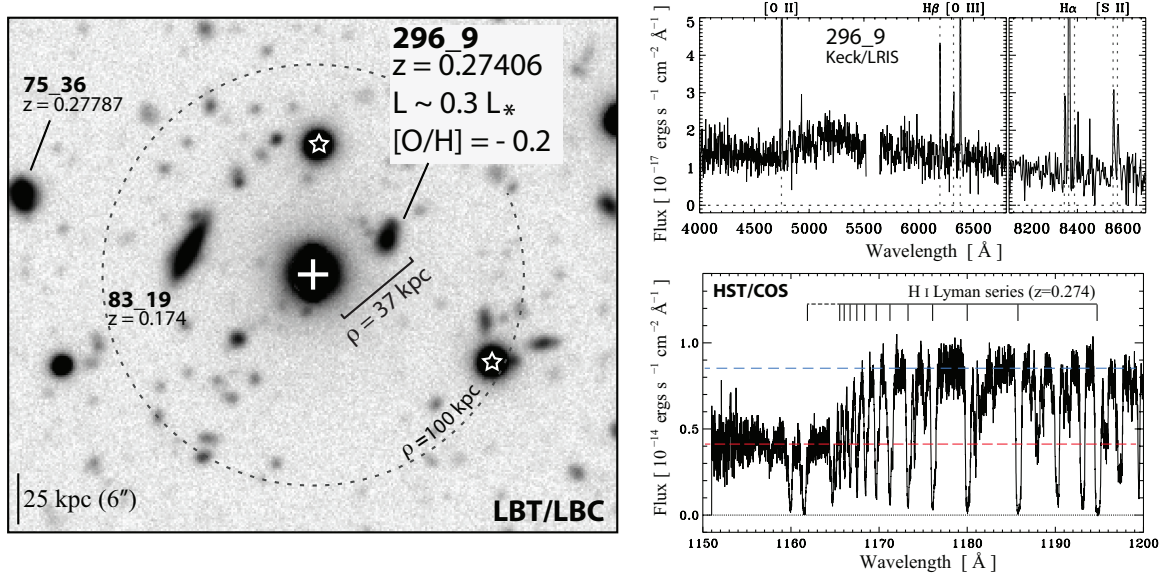}
\caption{A small section of a 2.5 ks LBT g-band image of the
PG1630+377 field (left). The dashed circle shows a 100 kpc radius ($\sim25 ''$) about the QSO 
 at $z = 0.274$. The host galaxy candidate is labeled 296\_9 and is a 
projected 37 kpc from the QSO line of sight (marked with a +). Other galaxies in the field  
with spectroscopic redshifts are noted. Foreground stars are marked with a star symbol. 
The LRIS spectrum of galaxy 296\_9 (upper right) 
shows the emission lines characteristic of a star forming galaxy. 
A portion of the COS spectrum of PG1630+377 
(lower right) shows the LLS located at $z=$\zone\ and the Lyman series lines leading 
up to the break. The \hi\ column density of the LLS ($\log N({\rm H\,I})=17.06\pm0.05$) was calculated by 
comparing the estimated unabsorbed continuum flux (upper dashed line) and the estimated 
mean absorbed flux (lower dashed line).}
\label{fig:big}
\end{figure*}


Recent surveys of LLSs have examined the statistical nature of the
absorber population, delineating the redshift evolution of these absorbers
\citep{prochaska10,sc10,ribaudo11}.  The literature contains only a
few LLSs for which the physical properties of the gas (metallicity,
ionization structure, kinematics) and the host galaxy (luminosity,
metallicity, mass) are well constrained
\citep{cp00,jenkins05,tripp05,prochaska06,cooksey08,lehner09,tumlinson11}.
To understand the implications for the statistical evolution of LLSs in
the context of galaxy evolution it is necessary to better understand
the gas-galaxy relationship, including the frequency with which LLSs
trace infall onto versus outflows from galaxies.

Here we use observations from the Cosmic Origins Spectrograph (COS),
with additional ground-based spectroscopic and imaging observations,
to analyze a LLS at $z\sim0.274$ along the sight line to the UV-bright
QSO PG1630+377 ($z_{\rm em}=1.476$).  We demonstrate the LLS has low
metallicity (\S~\ref{sec:absorption}) and use Large Binocular
Telescope (LBT) imaging and Keck spectroscopy of galaxies in the QSO
field to identify a near-solar metallicity, 0.3 L$_{*}$ galaxy at
virtually the same redshift (\S~\ref{sec:galenv}). We discuss this gas
in the context of cold mode accretion models in \S~\ref{sec:origins}
and summarize our results in \S~\ref{sec-sum}.


\section{Absorption From the Low Metallicity Lyman Limit System}
\label{sec:absorption}

\begin{figure*}[t]
\epsscale{0.75}
\plotone{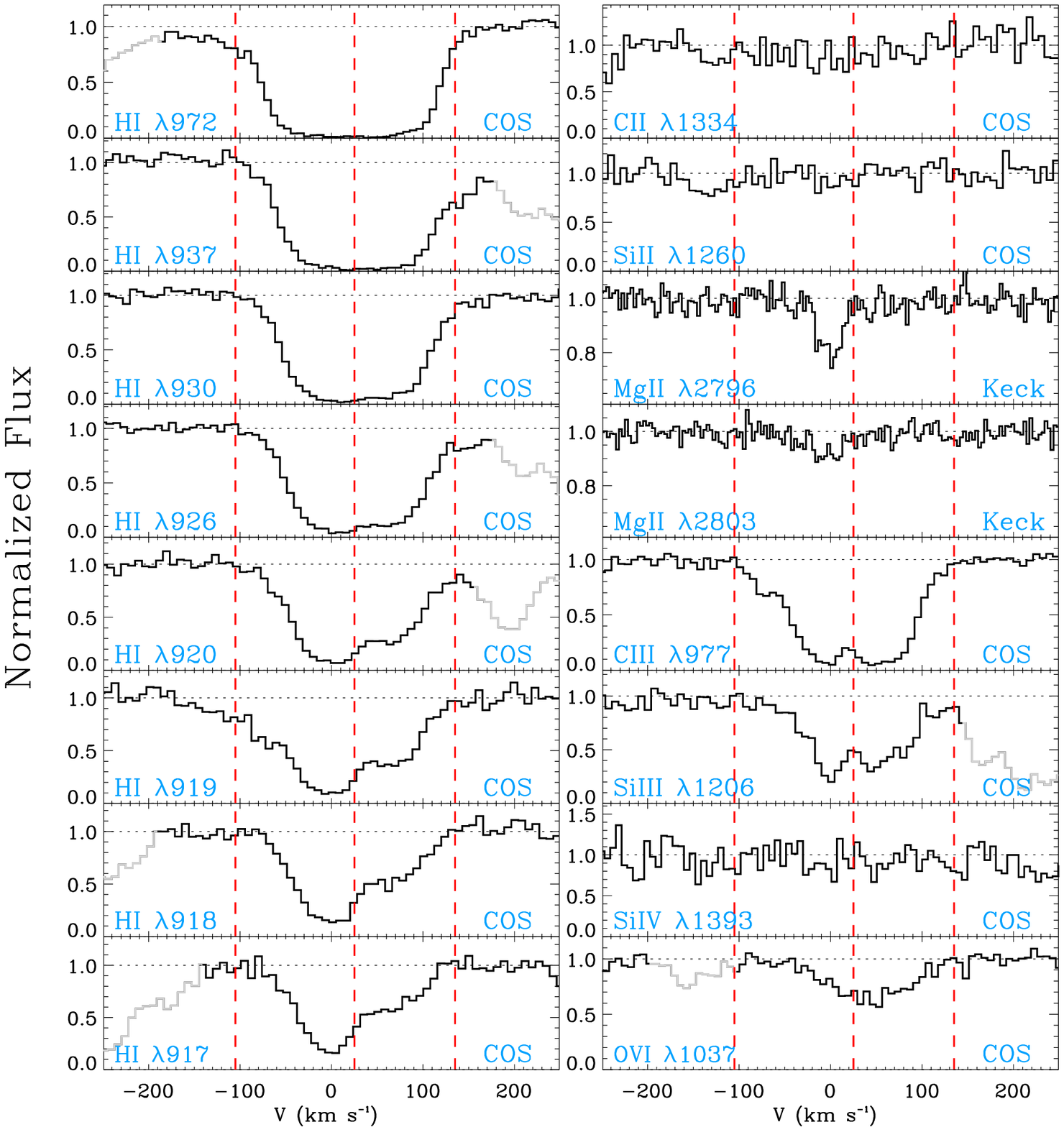}
\caption{Plots of absorption lines as a function of velocity centered on the 
LLS at $z=$\zone. The \hi\ Lyman series is shown on the left-hand side, while 
the right-hand side shows the metal ions. The dashed lines show the low ion 
integration limits adopted for the strong ([$-105,+25$] \kms) and weak components 
([$+25,+135$] \kms). We do not detect \cii, \silii, or \siliv\ in either component. 
The \ovi\ $\lambda$1031 is largely contaminated by another unrelated absorber and 
is not reproduced here. Gray portions of the spectra are absorption from unrelated 
absorbers. Note the scale of the panels showing the \mgii\ lines are set to a 
different scale than the rest of the panels.}
\label{fig:profile}
\end{figure*}

The UV spectra of PG1630+377 were obtained with COS on-board the {\em
  Hubble Space Telescope} (PID 11741, PI Tripp) using the G130M
(1150--1450 \AA) and G160M (1405--1775 \AA) gratings. The exposure
times for these two configurations were 23.0 and 14.3 ks,
respectively, giving S/N up to 30--40 per resolution element at
unabsorbed wavelengths. The excellent quality of the UV spectra is
displayed in Figures~\ref{fig:big} and \ref{fig:profile}.  The data
were processed using CALCOS (v2.11b) and coadded following
\citet{meiring11}. The lower-right panel of Figure~\ref{fig:big} shows
a small portion of the spectrum obtained from COS, highlighting the
presence of a LLS at $z\sim0.274$. Optical spectra were obtained with
HIRES at the Keck Observatory on 26 March 2010 and cover
\mgii\ $\lambda\lambda2796,2803$ at $z\sim0.274$. We acquired 3 HIRES
exposures totaling 2.2 ks under good conditions with the blue
cross-disperser, the C1 decker (0.86\arcsec\ width giving
FWHM~$\approx 6$ \kms), and the CCD mosaic.  The data were processed
using HIRedux in
IDL.\footnote{http://www.ucolick.org/$\sim$xavier/HIRedux/index.html}

\begin{figure*}[t]
\epsscale{0.75}
\plotone{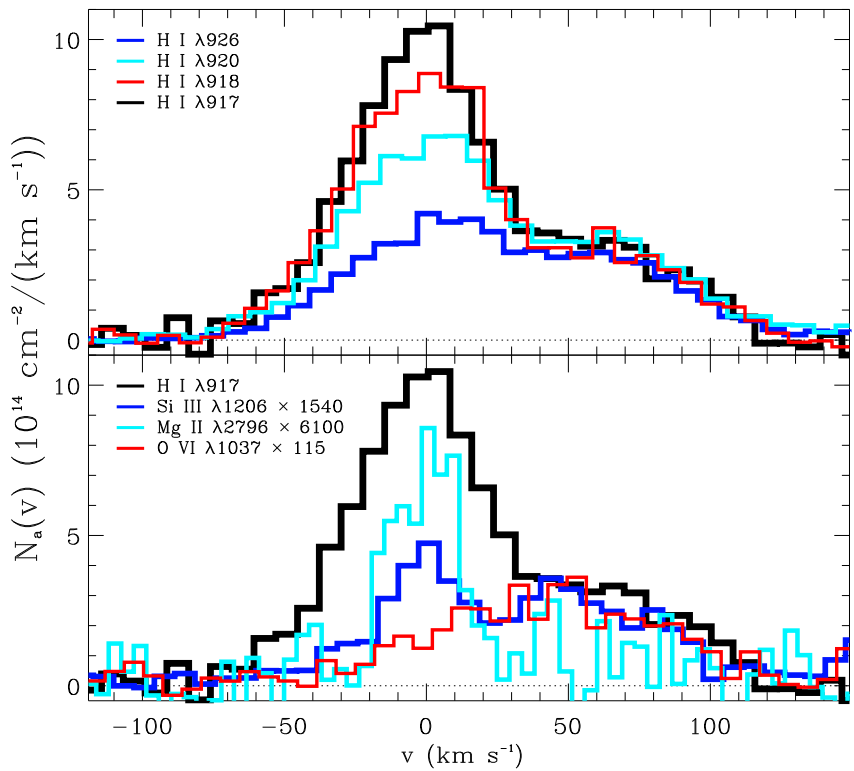}
\caption{The apparent column densities of several \hi\ transitions 
({\it top panel}) and \hi\ and metal ions ({\it bottom panel}). The top 
panel shows the different effects of saturation on the two main components 
of \hi, specifically the weak \hi\ component is unsaturated. In the bottom 
panel,  the apparent column densities of the ions have been scaled, revealing 
that (i) the \hi\ and weak/intermediate aligns well in the strong component, 
and (ii) the \ovi\ follows the lower ions and \hi\ of the weak component 
extremely well.}
\label{fig:nav}
\end{figure*}

Figure~\ref{fig:profile} shows the normalized absorption profiles of
\hi\ ({\it left panels}) and metal ions ({\it right panels}) as a
function of velocity relative to $z_{\rm abs}=\zone$, the centroid of
\mgii\ absorption. At $v=0$ \kms\ we detect \hi, \ciii, \siliii, weak
\mgii\ absorption ($W_r(2796)=59 \pm 4$ m\AA), and possibly \ovi.  We
refer to this as the strong component\footnote{Though we use the term
  components, the reader should be aware each is likely an unresolved
  blend of multiple sub-components or \textquotedblleft
  clouds.\textquotedblright} due to the strength of the \hi\
absorption.  Near $v=+50$ \kms\ ($z=\ztwo$) we detect \hi, \ciii,
\siliii, and \ovi, but no \mgii. We refer to this as the weak
component.  In Figure~\ref{fig:nav} we show the apparent column
density profiles, $N_a(v)$ \citep[][]{savage91}, of selected
species. This figure demonstrates that the \ovi\ follows the
intermediate ion \siliii\ and \hi\ in the weak component well and that
all of the observed \ovi\ could be due to the weak component.


The column densities and limits for selected species can be found in
Table~\ref{tab:stats}.  We determine the column densities of metal
ions by integrating their $N_a(v)$ profiles.  For non-detections, we
quote $3\sigma$ limits following \citet{lehner08}. From the flux
decrement at the Lyman limit, we measure the total optical depth of
the system to be $\tau_{\rm LLS} = 0.73 \pm 0.08$ (see
Figure~\ref{fig:big}), implying a total \hi\ column density $\log
N(\mbox{\hi})=17.06 \pm 0.05$ \citep[see][]{spitzer78}. We determine
the \hi\ column density of the weak component by integrating the
$N_a(v)$ profiles of the weak, unsaturated Lyman series lines. The top
panel in Figure~\ref{fig:nav} shows the constrast in the saturation
effect between the two components: for the weak component, the weak
Lyman series lines provide a good estimate of the \hi\ column
density. We find $\log N(\mbox{\hi})=16.30 \pm 0.02$ for the weak
component. The column density of the strong component is the
difference between the total and weak component column densities,
$\log N(\mbox{\hi})=16.98 \pm 0.06$. As a consistency check, we have
also fitted the \hi\ absorption profiles with a two component
model, wherein a model absorption profile is convolved with the COS
instrumental spread function to determine the best fit values of the
central velocities, column densities, and Doppler parameters of the
two assumed components.  The columns derived from such profile fitting
models are consistent with the integrated values, while the $b$-values
for both components are about 30 \kms.  We emphasize, however, that
the results from profile fitting depend upon the assumed component
structure of the gas.  While we have assumed a two component model,
each is quite likely made up of several blended absorbing components. 

To determine the metallicity and physical conditions of the strong
component, we model its ionization using Cloudy
\citep[v08.01,][]{ferland98}.  We assume the gas is photoionized,
modeling it as a uniform slab in thermal and ionization equilibrium.
We illuminate the slab with the \citet{hm11} 
background radiation field from quasars and
galaxies. We vary the ionization parameter, $U=n_\gamma/n_H$, and
metallicity of the gas (assuming solar relative abundances from
\citet{ags09}) to match the observed column density constraints
(Table~\ref{tab:stats}).

\begin{figure}
\epsscale{1.0}
\plotone{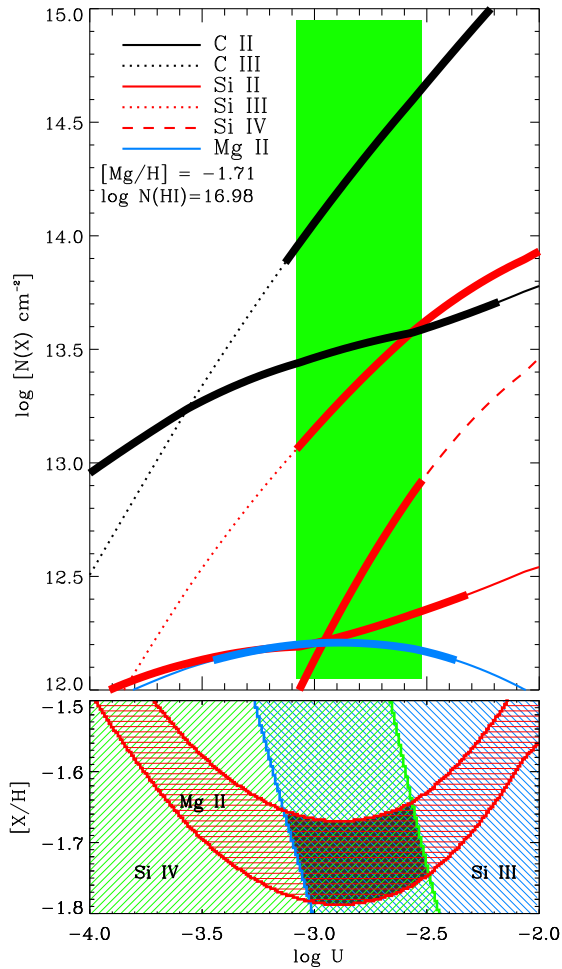}
\caption{The top panel shows the Cloudy-predicted column densities as a 
function of ionization parameter for the strong component with the 
metallicity of the gas set to [Mg/H]$=-1.71$. The bold portions of the 
curves show where the model column densities are consistent with the 
observations. The green band shows the range of ionization parameter 
for which the models are consistent with our observations. The 
lower panel shows the metallicities and ionization parameters consistent 
with the observed columns of the Si and Mg ions. The overlap of these near 
[X/H]$=-1.7$ and $\log U=-3$ to $-2.5$ is the region allowed by these constraints. 
The \mgii\ region includes the uncertainties in \hi\ since these directly 
affect the [X/H] measurements. The detection of \mgii\ absorption is
crucial for determining the metallicity to better than a factor of $\sim3$.}
\label{fig:cloudy}
\end{figure}

We summarize the results of the Cloudy simulations for the strong
component in Figure~\ref{fig:cloudy}. The ionization parameter is very
well constrained by the adjacent ionization states of Si and C, while
the metallicity of the gas is fixed mostly by \mgii. The observed
column densities are reproduced for models with an ionization
parameter $\log U = -2.80 \pm 0.30$, which is represented by the green
shaded region in the upper panel. This $\log U$ gives a metallicity of
$[\rm Mg/\rm H]=-1.71\pm 0.06$. For this range of $\log U$ the gas is
almost completely ionized, with a neutral hydrogen fraction, $X({\rm
  H\,I})=N(\rm HI)/N(\rm H)\sim 0.001- 0.004$. The particle
density in these models is $n_{H} \sim 0.001 - 0.003$ cm$^{-3}$, the total
H column is $\log N(\rm H) \sim 19.6$, the physical size of the
absorbing cloud is $L=N_{\rm H}/n_{H}\sim$2--25 kpc, and the
temperature of the gas is predicted to be $T\sim(2-4)\times 10^4$
K. This temperature is consistent with  the
component fitting where $b(HI)\sim30$ \kms\, which implies a temperature 
of $\sim3\times 10^4$ K.
 We note that the ionization state of this gas is unusually well
constrained by the lack of either \silii\ or \siliv\ absorption and
the presence of strong \siliii.


The intermediate and low ions in the weak component can be described
by similar photoionization models with a metallicity consistent with
the strong component.  However, this does not explain the strong
\ovi\ absorption.  We hypothesize the weak component is a multiphase structure
in which low-metallicity gas is interacting directly with the gaseous
corona of the host galaxy (\S~\ref{sec:galenv}).  In this scenario
the \ovi\ is produced in the interface
between the low-ionization gas and the corona, analagous to the
\ovi\ seen in the Galactic high velocity clouds \citep{sembach03}.
The close kinematic relationship of the intermediate and high ion
profiles (Figure \ref{fig:nav}) is consistent with this hypothesis \citep{howk09}.
The close kinematic relationship of the intermediate and high ion
profiles, however, could also suggest another possibility in 
which all the ions trace the same gas. The ionic ratios (\ciii/\ovi, \siliii/\ovi,
\siliii/\hi) can be matched by nonequilibrium models \citep{gs07} if
the metallicity of the gas is near solar.  In this case, the weak
component would represent radiatively-cooling material associated
with the host galaxy. The close correspondence between the \ovi\ and
the \hi\ in the weak component is not unusual; at low redshifts, \ovi\ and \hi\
absorption profiles are often observed to be well-aligned and to have similar 
profile shapes \citep{tripp08}.


\begin{deluxetable}{lcc}
\tabletypesize{\footnotesize}
\tablewidth{0pt} 
\tablecaption{PG1630+377 Absorber Properties \label{tab:stats}}
\tablehead{ 
\colhead{Species}& \multicolumn{2}{c}{$\log N$ [cm$^{-2}$]\tablenotemark{a}}\\
\cline{2-3} \\
\colhead{} & \colhead{Strong\tablenotemark{b}} & \colhead{Weak\tablenotemark{c}}
}
\startdata
%
\hi\ & $16.98 \pm 0.06$ & $16.30 \pm 0.02$ \\
\cii\ & $<13.71$ & $<13.50$ \\
\ciii\ & $>13.87$ & $>13.91$ \\
\silii\ & $<12.42$ & $<12.38$ \\
\siliii\ & $>13.07$ & $>13.11$ \\
\siliv\ & $<12.93$ & $<12.89$ \\
\mgii\ & $12.19 \pm 0.02$ & $<11.50$ \\
\ovi\ & \nodata & $14.48 \pm 0.03$\tablenotemark{d}\\
\enddata
\tablecomments{We adopt oscillator strengths from \citet{morton03}.}
\tablenotetext{a}{Strong and weak \hi\ components are integrated over $\Delta v=[-105,+25]$ \kms\ 
and $\Delta v=[+25,+135]$ \kms\ with respect to $z=$\zone.}
\tablenotetext{b}{Centered at $z=$\zone.}
\tablenotetext{c}{Centered at $z=$\ztwo.}
\tablenotetext{d}{This is all the detected \ovi, integrated over 
$\Delta v=[-35,+135]$ \kms. \ovi\ in the strong component velocity range is $\log N(\rm O VI)=13.91 \pm 0.04$.}
\end{deluxetable}



\section{The Galactic Environment of the Lyman Limit System}
\label{sec:galenv}

The left panel of Figure~\ref{fig:big} shows a $g$-band image of the
field toward PG1630+377 obtained with the Large Binocular Camera (LBC)
on the LBT.  The 2.5 ks LBC image was taken with $\sim1\arcsec$ seeing
and reaches 5$\sigma$ limiting magnitude of $g_{AB}\sim25.5$ mag in a
$2\arcsec$ aperture, equivalent to $L\sim0.005 \, L_*$ at
$z\sim0.274$.  A full description of the LBT imaging will be presented
in a future paper.

We selected targets for a preliminary redshift survey based on
photometric redshifts of bright galaxies from the Sloan Digital Sky
Survey (SDSS).  For the three galaxies marked in Figure~\ref{fig:big},
we obtained longslit spectra with the Keck/LRIS through a $1\arcsec$
slit using the D560 dichroic with the 600/4000 grism (blue side) and
600/7500 grating (red side).  This gave spectral coverage between 3000
to 5500 \AA\ (blue side) and 5600 to 8200 \AA\ (red side).  On the
blue side, binning the data 2 $\times$ 2 resulted in a dispersion of
1.2 \AA\ per pixel and a resolution of R$\sim1070$ (FWHM$\sim$280 \kms). On the
red side, the data were binned 2 $\times$ 1, resulting in a dispersion
of 2.3 \AA\ per pixel and a resolution of $\sim$200 \kms.  For strong
nebular emission lines from the galaxies we achieved S/N$\ge10$
pixel$^{-1}$ with exposure times of $1\times900$ s in the blue and
$2\times410$ s in the red.  

The data reduction and calibration \citep[see][]{werk11} were
carried out using the
LowRedux\footnote{http://www.ucolick.org/$\sim$xavier/LowRedux/index.html}
IDL software package, which includes flat fielding to correct for
pixel-to-pixel response variations and larger scale illumination
variations, wavelength calibration, sky subtraction, and flux
calibration using the spectrophotometric standard star G191-B2B.  We
also applied a flux correction to the spectra using SDSS photometry to
ensure the flux calibrations of the red and blue side spectra were
consistent and to correct for light losses in the 1\arcsec\ slit.  To
do this we convolved the LRIS spectra with SDSS $ugriz$ filter response
curves using the IDL code ``spec2mag'' \citep[see][]{silva11}.  
We then compared the spectrally-determined apparent
magnitudes with the SDSS catalog apparent magnitudes \citep[see][]{werk11} 
to derive the correction.  We corrected the
spectra for foreground Galactic reddening using the maps of
\citet{schlegel98} and assume an intrinsic ratio of H$\alpha$ to
H$\beta$ of 2.86 to correct for internal reddening, which corresponds
to case B recombination at an effective temperature of 10,000 K and
electron density of 100 cm $^{-3}$ \citep{hs87}.

We determined the galaxy redshifts by fitting template spectra to the
LRIS spectra using a modified version of the SDSS {\it zfind} code.
Systematic effects (e.g., wavelength calibration) dominate the errors,
limiting the accuracy of our measurements to $\sim25$ \kms.  We
eliminate one of the candidate hosts (83\_19)\footnote{The first
  number in this notation gives the direction in degrees of the galaxy
  north of east from this QSO line of sight. The second number is the
  angular distance in arcseconds to the galaxy from the line of
  sight.} due to its low redshift ($z\sim0.174$).  The other two
galaxies have redshifts similar to that of the LLS (see
Figure~\ref{fig:big}) at impact parameters\footnote{In this paper we
  assume $H_0=72$ \kms\ Mpc$^{-1}$, $\Omega_m=0.27$, and
  $\Omega_{\Lambda}=0.73$.} of $37$ kpc (296\_9) and $140$ kpc
(75\_{36}).  However, the velocity of 75\_36 relative to the absorber
is likely too high for it to be the associated galaxy: $\Delta
v_{75\_{36}} \equiv v_{\rm abs} - v_{\rm gal} = -925$ \kms .  The
velocity separation between 296\_9 and the LLS is $\Delta v_{296\_9} =
-26$ \kms. We identify this galaxy as associated with the LLS.  There
are a number of other galaxies in the field without spectroscopic
redshifts.  However, the identification of the LLS with 296\_9 is
secure for the following reasons.  Most importantly, the velocity
separation between the absorber and this galaxy is extremely small.
It is very unlikely for such a small velocity offset to occur
randomly.  At the very least, the absorber is associated with the
environment with which 269\_9 is associated.  In addition, all of the
galaxies nearer to the QSO sight line than the candidate host are
extremely faint with very small angular diameters, consistent with
their being high redshift background systems.

We use the observed H$\alpha$ luminosity of 269\_9 with the
\citet{calzetti10} relationship to estimate a star formation rate of
this galaxy, deriving ${\rm SFR}_{\rm 296\_9} = 1.5$ M$_{\odot}$
yr$^{-1}$.  The stellar mass of 296\_9 is $M_\star\sim2\times10^9
M_\odot$, derived from SDSS photometry.  We determined the oxygen
abundance using the R23 metallicity indicator \citep{pagel79} as
calibrated by \citet{mcgaugh91}. We derive an abundance for 296\_9 of
$[\rm O/\rm H]=-0.20\pm0.15$. This estimate places the object near the
convergence of the two branches of the R23 diagnostic, thus the
systematic errors due to the degeneracy in metallicities associated
with the strong line method are small compared to the factor of 30
difference in metallicities seen between the LLS gas and the
galaxy. We used the [\nii]/[\oii] ratio, following \citet{ke08}, to
confirm the derived abundance.  The general systematic errors
associated with age effects and stellar distributions can be as large
as 0.25 dex \citep{ercolano07}, but such large effects are still not
enough to modify our conclusions. The metallicity, stellar mass, and star 
formation rate derived for 296\_9 place this system on the mean mass-metallicity
relationship of \citet{mannucci10}.


\section{The Origin of Primitive Gas in Lyman Limit Systems}
\label{sec:origins}

We have identified a low-redshift, $z_{\rm abs}\sim 0.274$, LLS with a
very low metallicity, $[\rm Mg/\rm H]=-1.71 \pm 0.06$, associated with
a near-solar metallicity galaxy.
The metallicity of the LLS is much lower than the metallicity
distribution of damped Lyman-$\alpha$ systems (DLAs) at $z \la 1$ 
\citep{kf02,rao05,wolfe05}. From a sample of 11 DLAs at $z \la 1$ 
\citep[see Table 5,][]{rao05}, the mean metallicity of these systems is
around $-0.7$ dex; these systems have a
large spread in metallicity, but the two lowest values are $-1.2$ and
$-1.3$ dex. 
In contrast, such low metallicities do not appear to be the
exception for LLSs at $z\la 1$.  A perusal of the literature finds 4
out of 8 well-studied LLS at $z\la1$ have metallicities $[{\rm M/H}]
<-1.8$ to $\approx -1.4$ \citep[this
work,][]{zonak04,tripp05,cooksey08}. Recently, \citet{thom11} found
another strong \hi\ absorber with $\log N(\mbox{\hi}) \sim 16$ having
$[{\rm M/H}]<-1.5$. Preliminary results of our on-going study to
expand the number of LLS metallicity measurements at $z \la 1$ appear
to confirm that the metallicity distribution of LLSs is different from
that of DLAs at similar redshifts, with a high fraction of LLSs having
$[{\rm M/H}]\la -1.5$ (N. Lehner et al. 2011, in prep.). 

\citet{zonak04} and \citet{tripp05} argued that a possible origin for these
low metallicity LLSs was low metallicity dwarf galaxies. However, the
(tentative) high frequency of these primitive LLSs could favor another
origin that was not really considered previously. Indeed, since only the tail of
the dwarf galaxy metallicity distribution shows such very low abundances,
they are unlikely to show up in significant numbers in absorption line
studies.  That is, the probability a line of sight passing through gas
from galaxies like the very low metallicity dwarf galaxy I\,Zw\,18
\citep{ks86} is very small. Furthermore, the relatively low \hi\ column 
density of the system demonstrates the sight line does not pass through
a galaxy.
Except for the highest-redshift LLSs in this sample
\citep{zonak04}, all of these LLSs show at least a sub-$L*$
galaxy within 100 kpc consistent with being a ``host'' galaxy.  For
the present sight line, our deep LBC/LBT images of the field do not
show other strong candidates for a galaxy associated with the LLS at
$\rho \la 37$ kpc, though our spectroscopic galaxy information is
limited.

For the present LLS, we have demonstrated that the
likely host galaxy has a metallicity much higher than the LLS, ruling
out altogether that the gas in the LLS could originate from this
galaxy.  We therefore propose instead that this and the
low-metallicity LLSs from the literature trace matter infalling onto
galaxies, perhaps related to the CMA streams predicted by galaxy
formation simulations.  The physical properties of the low-metallicity
LLSs are similar to those predicted by CMA models, including the
temperature, metallicity, and \hi\ column as well as host galaxy
properties such as velocity offset and mass.  For the present LLS, the
temperature of this LLS is constrained to be $T\le 3.8\times10^4$ K
from the $b$-value derived from our component fit to the \hi\
profiles, consistent with the temperatures predicted by all CMA
models. More generally, these LLSs are well modeled by photoionization
models that predict temperatures to be about a few times $10^4$ K.

The low metallicity of the present and literature systems are
consistent with the predictions of \citet{fumagalli11}, who find cold
streams in their simulations have a broad distribution of abundances
centered on $\sim1\%$ solar. 
These LLSs are likely to have been enriched above primordial
levels by previous star formation episodes before streaming into a
galaxy. Fumagalli et al. predict the majority of cold streams should
be predominantly ionized at $\log N({\rm H\,I})\la 19.0$, consistent
with the properties of the current sample of LLSs. For our study, we
estimate that the stellar mass of the host galaxy is
$M_\star=2\times10^9 M_\odot$, well below the predicted stellar mass
cutoff for CMA host galaxies.  While there could always be galaxies
hidden underneath the glare of the background QSO that could have
contributed the gas making up this LLS, we note that this is also
consistent with the CMA simulations.  The infalling streams in such
models are enriched by small galaxies that will themselves merge with
the dominant host galaxy.  

Our work highlights the potential importance of \hi-selected LLSs as
probes of infalling, metal-poor gas. While observations of some LLSs
clearly indicate they are related to galactic outflows/winds or galaxy
mergers \citep{jenkins05,lehner09,tumlinson11}, it is also
apparent that the population of LLSs includes very low-metallicity gas with
properties very similar to those predicted by CMA models. While the
sample of LLSs is still currently small, we will expand it at $z\la 1$
in the near future with the aim to better constrain the metallicity
distribution of these systems. Our analysis suggests the best metal
species for tracing CMA may be intermediate ions such as \ciii\ and
\siliii\ (and other ions like \oii\ or \oiii), as the low ions (\cii,
\silii, and \mgii) and high ions ( \siliv) 
are supressed in such low density, low metallicity
gas. \citet{kacprzak10} have also argued, on the basis of numerical
simulations, that the majority of weak \mgii\ systems ($20$ m\AA\ $<
W_r(2796) < 300$ m\AA) trace material infalling onto a galaxy. The
present LLS has $W_r(2796)=59 \pm 4$ m\AA, while the metal-poor system
studied by Zonak et al. (2004) has $W_r(2796)=97 \pm 8$ m\AA\ -- both
consistent with this suggestion.


\section{Summary}\label{sec-sum}

We have presented high-quality UV and optical spectroscopic and
imaging observations from COS/{\it HST}, HIRES/Keck, LRIS/Keck,and
LBC/LBT of a LLS with $\log N({\rm H\,I})=17.06\pm0.05$ at $z_{\rm
  abs} = \zone$ along the QSO PG1630+377 and the galaxies in its field
of view in order to explore the origin(s) of the LLS at low
redshift. The main results of our analysis are as follows:

\begin{enumerate}

\item The LLS shows metal line absorption from \mgii, \ciii,
  \siliii, and \ovi. The high S/N of the COS spectrum allowed us
  to derive stringent upper limits for several other
  ions (in particular \cii, \silii, and \siliv).  These allow a secure
  metallicity measurement of $[\rm Mg/\rm H] =-1.71 \pm 0.06$ for the
  stronger of the two components in this system due to the tight
  constraints on the ionization state of the gas.  The
  \ovi\ absorption is mostly associated with the weaker component (at $+50$
  \kms\ from the stronger absorber) and is likely a result of its
  interaction with a galaxy halo.

\item Our limited redshift survey of galaxies close on the sky to
  PG1630+377 shows that there is a 0.3 $L_*$ star-forming galaxy
  projected 37 kpc from the QSO at nearly identical redshift to the
  LLS ($z=\zgone, \Delta v = -26$ \kms) with near solar metallicity
  ($[\rm O/\rm H]=\metal$). While our spectroscopic survey is
  extremely limited, the deep LBT images do not suggest another
  plausible host galaxy at $\rho < 37$ kpc.

\item The literature contains several other very low metallicity LLSs
  at $z\la 1$. While the sample is still small, we find a high
  frequency ($\sim$50\%) of $z\la 1$ LLSs have $[{\rm M/H}] \la -1.5$,
  which appears to differ from the metallicity distribution of DLAs at
  similar redshifts and from a similar sample size. 
  We propose these very low metallicity LLSs are
  signatures of infall onto galaxies, perhaps similar to the cold mode
  accretion predicted by cosmological simulations.  The ionization,
  metallicity, temperature, and host galaxy properties in these
  simulations are in good agreement with the properties derived from
  observations of the \hi-selected LLSs.
 
\end{enumerate}

\acknowledgments Support for Program number HST-GO-11741 was provided
by NASA through a grant from the Space Telescope Science Institute,
which is operated by the Association of Universities for Research in
Astronomy, Incorporated, under NASA contract NAS5-26555.  JCH and JR
acknowledge support from NASA grant NNX08AJ31G. TMT and JDM also
appreciate support from NASA grant NNX08AJ44G.
This work made use of data from the Large Binocular Telescope.  The
LBT is an international collaboration among institutions in the United
States, Italy and Germany. The LBT Corporation partners are: The
University of Arizona on behalf of the Arizona university system;
Istituto Nazionale di Astrofisica, Italy; LBT
Beteiligungsgesellschaft, Germany, representing the Max Planck
Society, the Astrophysical Institute Potsdam, and Heidelberg
University; The Ohio State University; The Research Corporation, on
behalf of The University of Notre Dame, University of Minnesota and
University of Virginia.



\begin{thebibliography}{}

\bibitem[Asplund et 
al.(2009)]{ags09} Asplund, M., Grevesse, N., Sauval, A.~J., \& Scott, P.\ 2009, \araa, 47, 481

\bibitem[Bauermeister et al.(2010)]{bauermeister10} Bauermeister, A., 
Blitz, L., \& Ma, C.-P.\ 2010, \apj, 717, 323

\bibitem[Birnboim 
\& Dekel(2003)]{bd03} Birnboim, Y., \& Dekel, A.\ 2003, \mnras, 345, 349

\bibitem[Bouch{\'e} et al.(2007)]{bouche07} Bouch{\'e}, N., et 
al.\ 2007, \apj, 671, 303

\bibitem[Bowen 
\& Chelouche(2011)]{bc11} Bowen, D.~V., \& Chelouche, D.\ 2011, \apj, 727, 47

\bibitem[Calzetti et al.(2010)]{calzetti10} Calzetti, D., et al.\ 
2010, \apj, 714, 1256

\bibitem[Chen 
\& Prochaska(2000)]{cp00} Chen, H.-W., \& Prochaska, J.~X.\ 2000, \apjl, 543, L9

\bibitem[Chen et al.(2010)]{chen10} Chen, H.-W., Wild, V., 
Tinker, J.~L., Gauthier, J.-R., Helsby, J.~E., Shectman, S.~A., 
\& Thompson, I.~B.\ 2010, \apjl, 724, L176

\bibitem[Cooksey et al.(2008)]{cooksey08} Cooksey, K.~L., 
Prochaska, J.~X., Chen, H.-W., Mulchaey, J.~S., 
\& Weiner, B.~J.\ 2008, \apj, 676, 262

\bibitem[da Silva et al.(2011)]{silva11} da Silva, R.~L., 
Prochaska, J.~X., Rosario, D., Tumlinson, J., 
\& Tripp, T.~M.\ 2011, \apj, 735, 54

\bibitem[Ercolano et al.(2007)]{ercolano07} 
Ercolano, B., Bastian, N., \& Stasi\'{n}ska, G. 2007, MNRAS, 379, 945

\bibitem[Faucher-Gigu{\`e}re 
\& Kere{\v s}(2011)]{fgk11} Faucher-Gigu{\`e}re, C.-A., \& Kere{\v s}, D.\ 2011, \mnras, L208

\bibitem[Ferland et al.(1998)]{ferland98} Ferland, G.~J., 
Korista, K.~T., Verner, D.~A., Ferguson, J.~W., Kingdon, J.~B., 
\& Verner, E.~M.\ 1998, \pasp, 110, 761

\bibitem[Fumagalli et al.(2011)]{fumagalli11} Fumagalli, M., 
Prochaska, J.~X., Kasen, D., Dekel, A., Ceverino, D., 
\& Primack, J.~R.\ 2011, arXiv:1103.2130

\bibitem[Gnat 
\& Sternberg(2007)]{gs07} Gnat, O., \& Sternberg, A.\ 2007, \apjs, 168, 213

\bibitem[Haardt 
\& Madau(2011)]{hm11} Haardt, F., \& Madau, P.\ 2011, arXiv:1105.2039

\bibitem[Howk et al.(2009)]{howk09} Howk, J.~C., Ribaudo, 
J.~S., Lehner, N., Prochaska, J.~X., 
\& Chen, H.-W.\ 2009, \mnras, 396, 1875

\bibitem[Hummer 
\& Storey(1987)]{hs87} Hummer, D.~G., \& Storey, P.~J.\ 1987, \mnras, 224, 801

\bibitem[Jenkins et al.(2005)]{jenkins05} Jenkins, E.~B., Bowen, 
D.~V., Tripp, T.~M., \& Sembach, K.~R.\ 2005, \apj, 623, 767

\bibitem[Kacprzak et al.(2010)]{kacprzak10} Kacprzak, G.~G., 
Churchill, C.~W., Ceverino, D., Steidel, C.~C., Klypin, A., 
\& Murphy, M.~T.\ 2010, \apj, 711, 533

\bibitem[Kacprzak et al.(2011)]{kacprzak11} Kacprzak, G.~G., 
Churchill, C.~W., Barton, E.~J., \& Cooke, J.\ 2011, \apj, 733, 105

\bibitem[Kere{\v s} et al.(2005)]{keres05} Kere{\v s}, D., 
Katz, N., Weinberg, D.~H., \& Dav{\'e}, R.\ 2005, \mnras, 363, 2

\bibitem[Kewley 
\& Ellison(2008)]{ke08} Kewley, L.~J., \& Ellison, S.~L.\ 2008, \apj, 681, 1183

\bibitem[Kulkarni 
\& Fall(2002)]{kf02} Kulkarni, V.~P., \& Fall, S.~M.\ 2002, \apj, 580, 732 

\bibitem[Kunth 
\& Sargent(1986)]{ks86} Kunth, D., \& Sargent, W.~L.~W.\ 1986, \apj, 300, 496

\bibitem[Lehner et al.(2008)]{lehner08} Lehner, N., Howk, J.~C., 
Keenan, F.~P., \& Smoker, J.~V.\ 2008, \apj, 678, 219

\bibitem[Lehner et al.(2009)]{lehner09} Lehner, N., Prochaska, 
J.~X., Kobulnicky, H.~A., Cooksey, K.~L., Howk, J.~C., Williger, G.~M., 
\& Cales, S.~L.\ 2009, \apj, 694, 734 

\bibitem[Mannucci et al.(2010)]{mannucci10} Mannucci, F., Cresci, 
G., Maiolino, R., Marconi, A., \& Gnerucci, A.\ 2010, \mnras, 408, 2115

\bibitem[McGaugh(1991)]{mcgaugh91} McGaugh, S.~S.\ 1991, \apj, 
380, 140

\bibitem[Meiring et al.(2011)]{meiring11} Meiring, J.~D., et al.\ 
2011, \apj, 732, 35

\bibitem[M{\'e}nard 
\& Chelouche(2009)]{mc09} M{\'e}nard, B., \& Chelouche, D.\ 2009, \mnras, 393, 808

\bibitem[Morton(2003)]{morton03} Morton, D.~C.\ 2003, \apjs, 
149, 205

\bibitem[Pagel et al.(1979)]{pagel79} Pagel, B.~E.~J., Edmunds, 
M.~G., Blackwell, D.~E., Chun, M.~S., \& Smith, G.\ 1979, \mnras, 189, 95

\bibitem[Prochaska et al.(2006)]{prochaska06} Prochaska, J.~X., 
Weiner, B.~J., Chen, H.-W., \& Mulchaey, J.~S.\ 2006, \apj, 643, 680

\bibitem[Prochaska et al.(2010)]{prochaska10} Prochaska, J.~X., 
O'Meara, J.~M., \& Worseck, G.\ 2010, \apj, 718, 392

\bibitem[Rao et al.(2005)]{rao05} Rao, S.~M., Prochaska, 
J.~X., Howk, J.~C., \& Wolfe, A.~M.\ 2005, \aj, 129, 9

\bibitem[Ribaudo et al.(2011)]{ribaudo11} Ribaudo, J., Lehner, 
N., \& Howk, J.~C.\ 2011, \apj, 736, 42

\bibitem[Savage \& Sembach(1991)]{savage91} Savage, B.~D., \& Sembach, K.~R.\ 1991, \apj, 379, 245

\bibitem[Schlegel et al.(1998)]{schlegel98} Schlegel, D.~J., 
Finkbeiner, D.~P., \& Davis, M.\ 1998, \apj, 500, 525

\bibitem[Sembach et al.(2003)]{sembach03} Sembach, K.~R., et al.\ 
2003, \apjs, 146, 165

\bibitem[Songaila \& Cowie(2010)]{sc10} Songaila, A., \& Cowie, L.~L.\ 2010, \apj, 721, 1448

\bibitem[Spitzer(1978)]{spitzer78} Spitzer, L.\ 1978, New York 
Wiley-Interscience, Physical Processes in the Interstellar Medium, 1978.~333 p.

\bibitem[Stewart et al.(2010)]{stewart10} Stewart, K.~R., 
Kaufmann, T., Bullock, J.~S., Barton, E.~J., Maller, A.~H., Diemand, J., 
\& Wadsley, J.\ 2010, arXiv:1012.2128

\bibitem[Thom et al.(2011)]{thom11} Thom, C., Werk, J.~K., 
Tumlinson, J., Prochaska, J.~X., Meiring, J.~D., Tripp, T.~M., 
\& Sembach, K.~R.\ 2011, arXiv:1105.4601

\bibitem[Tripp et al.(2005)]{tripp05} Tripp, T.~M., Jenkins, 
E.~B., Bowen, D.~V., Prochaska, J.~X., Aracil, B., 
\& Ganguly, R.\ 2005, \apj, 619, 714

\bibitem[Tripp et al.(2008)]{tripp08} Tripp, T.~M., Sembach, 
K.~R., Bowen, D.~V., Savage, B.~D., Jenkins, E.~B., Lehner, N., 
\& Richter, P.\ 2008, \apjs, 177, 39

\bibitem[Tumlinson et al.(2011)]{tumlinson11} Tumlinson, J., et 
al.\ 2011, \apj, 733, 111

\bibitem[Werk et al.(2011)]{werk11} Werk, J.~K., Prochaska, 
J.~X., Thom, C., et al.\ 2011, arXiv:1108.3852

\bibitem[Wolfe et al.(2005)]{wolfe05} Wolfe, A.M., Gawiser, E.,
  Prochaska, J.X. 2005, \araa, 43, 861

\bibitem[Zonak et al.(2004)]{zonak04} Zonak, S.~G., Charlton, 
J.~C., Ding, J., \& Churchill, C.~W.\ 2004, \apj, 606, 196

\end{thebibliography}
\end{document}